\documentclass[showkeys,aps,amsmath,amssymb,twocolumn,longbibliography,titlepage]{revtex4-1}
\usepackage{amssymb,amsmath,amsfonts}
\usepackage{bm}
\usepackage{graphicx}
\usepackage{hyperref}
\usepackage{times}

\usepackage{xcolor,soul}

\begin{document}

\title{Observation of the density dependence of the closed-channel fraction of a  $^6$Li superfluid}

\author{Xiang-Pei Liu$^{1,2,3}$}
\thanks {X.-P. Liu and X.-C. Yao contributed equally to this work.}
\author{Xing-Can Yao$^{1,2,3}$}
\thanks {X.-P. Liu and X.-C. Yao contributed equally to this work.}
\author{Hao-Ze Chen$^{1,2,3}$}
\author{Xiao-Qiong Wang$^{1,2,3}$}
\author{Yu-Xuan Wang$^{1,2,3}$}
\author{Yu-Ao Chen$^{1,2,3}$}
\email{yuaochen@ustc.edu.cn}
\author{Qijin Chen$^{1,2,3}$}
\email{qchen@uchicago.edu}
\author{K. Levin$^{4}$}
\author{Jian-Wei Pan$^{1,2,3}$}
\email{pan@ustc.edu.cn}
\affiliation{$^1$Hefei National Laboratory for Physical Sciences at the Microscale and Department of Modern Physics, University of Science and Technology of China, Hefei 230026, China}
\affiliation{$^2$Shanghai Branch, CAS Center for Excellence in Quantum Information and Quantum Physics, University of Science and Technology of China, Shanghai 201315, China}
\affiliation{$^3$Shanghai Research Center for Quantum Sciences, Shanghai 201315, China}
\affiliation{$^4$James Franck Institute, University of Chicago, Chicago, Illinois 60637, USA}

\begin{abstract}
  Atomic Fermi gases provide an ideal platform for studying the
  pairing and superfluid physics, using a Feshbach resonance between
  closed channel molecular states and open channel scattering
  states. Of particular interest is the strongly interacting
  regime. We show that the closed-channel fraction $Z_{cc}$ provides an
  effective probe for the important many-body interacting effects,
  especially through its density dependence, which is absent from
  two-body theoretical predictions.  Here we measure $Z_{cc}$ as a function
  of interaction strength and the Fermi temperature $T_\text{F}$ in a
  trapped $^6$Li superfluid throughout the entire BCS--BEC crossover,
  in quantitative agreement with theory when  important thermal
  contributions outside the superfluid core are taken into account.  Away from
  the deep BEC regime, the fraction $Z_{cc}$ is sensitive to
  $T_\text{F}$. In particular, our data show
  $Z_{cc} \propto T_\text{F}^{\alpha}$ with $\alpha=1/2$ at unitarity, in
  quantitative agreement with calculations of a two-channel pairing
  fluctuation theory, and $\alpha$ increases rapidly into the BCS
  regime, reflecting many-body interaction effects as predicted.

\end{abstract}

\maketitle


\section{Introduction}
Using a Feshbach resonance (FR), atomic Fermi gases~\cite{Bloch2008RoMP} provide an ideal
platform for studying the pairing and superfluid physics. Of particular interest is the crossover~\cite{Leggett,PhysRepReview} from
a Bardeen-Cooper-Schrieffer (BCS) state of Cooper pairs to a Bose-Einstein condensate (BEC) of molecular dimers~\cite{Jochim2003S,Greiner2003N,Kinast2005S,Navon2010S,Nascimbene2010N,Ku2014PRL, Ries2015PRL, Boettcher2016PRL}. Unlike in a one-channel model, where the effect of FR is oversimplified into a tunable non-retarded pairing interaction strength, the rich and important physics of FR can be described by a two-channel model~\cite{PhysRepReview}, in which open-channel atom pairs are linearly superimposed with closed-channel molecules~\cite{Duine2003PRL,Duine2004PR,Romans2005PRL}, and are further dressed with many-body interactions. These ``dressed molecules'' can be represented symbolically as
$ \Psi_\text{dressed} = \sqrt{ Z}\Psi_\text{closed} +\sqrt{1-
  Z}\Psi_\text{open} $,
where $Z$ reflects the closed-channel fraction within the ``dressed molecules''. While the superposition reflects two-body physics, the underlying dressing reflects important many-body physics.
Indeed, many important physical quantities, such as the superfluid excitation gap~\cite{Grimm2004S,Chen2005PRL,Jin2008N} and Tan's contact~\cite{Tan2008,Werner2009TEPJB,Sagi2012PRL,Hoinka2013PRL,Mukherjee2019PRL}, can be associated with closed-channel fraction. The measurement of how closed-channel fraction evolves with interaction, density, and temperature thus can provide crucial information of the many-body interaction effects and serve as a benchmark to test various many-body theories.

Experimentally~\cite{Partridge2005PRL}, it is more convenient to measure the closed-channel fraction $Z_{cc}$ of the entire Fermi gas. One could obtain ${Z}$ by dividing $Z_{cc}$ with the pair fraction \cite{PhysRepReview}, which is unity in the deep BEC regime but becomes small in the BCS regime. In the simple two-body theory~\cite{Chin2010RMP}, $Z_{cc}$ decreases from 1 in the BEC limit to 0 at unitarity, beyond which the attractive interaction becomes too weak to support bound molecules. When it is large, $Z_{cc}$ can be determined by the derivative of the binding energy of dressed molecules with respect to the magnetic field $B$~\cite{Wu2012PRL}. However, this method does not work when $Z_{cc}$ is small, due to limited experimental resolution as well as the breakdown of the two-body theory in the unitary and BCS regimes. By driving transitions between the dressed molecules and molecules in excited states with a resonant laser, $Z_{cc}$ has been previously measured in the BCS--BEC crossover of a $^6$Li superfluid~\cite{Partridge2005PRL}. While $Z_{cc}$ decreased from the BEC to the unitary regime, a non-vanishing and smoothly varying $Z_{cc}$ was observed across unitarity into the BCS regime~\cite{Partridge2005PRL}. Despite that the quantity $Z_{cc}$ was extracted assuming an exponential decay of remaining atom number $N$ versus laser probing time $t$ based on the two-body theory, this observation indicates that many-body effect must be present in the BCS regime. Indeed, $Z_{cc}$ was contemplated to be related to the square of excitation gap in this regime and slight deviation from exponential loss for $N$ was noticed~\cite{Partridge2005PRL}. Nonetheless, due to limited signal-to-noise ratio and large error bars in atomic numbers, a clear dependence of $Z_{cc}$ on the particle number $N$, which manifests the many-body effects, was not observed.

This experiment has been addressed to various degrees by many-body based two-channel models~\cite{Chen2005PRL,Romans2005PRL,Werner2009TEPJB,ZhangSZ2009}. It is shown that $Z_{cc}$ depends not only on the scattering length but also on the Fermi temperature $T_\text{F}$ of the system~\cite{Chen2005PRL,ZhangSZ2009}, unlike that in previous experiment~\cite{Partridge2005PRL}, which reported a single unique value of $Z_{cc}$ for given scattering length. Particularly, a universal relation of $Z_{cc}\propto \sqrt{T_\text{F}}$ at unitarity is predicted. A direct consequence of these predictions is that, although very weak, $Z_{cc}$ is a function of the particle number $N$, via $Z_{cc}\propto T_\text{F}^\alpha \propto N^{\alpha/3} $ with
$\alpha = 1/2$ at unitarity and $\alpha > 1/2$ in the BCS regime,
resulting in a power-law decay of atom number as a function of laser
probe time. Importantly, in the $N\rightarrow 0$ limit, $Z_{cc}$
vanishes both at unitarity and in the BCS regime, consistent with the
two-body result.

In this paper, we report on precision measurements of the
closed-channel fraction $Z_{cc}$ as a function of magnetic field and Fermi
temperature $T_\text{F}$ in $^6$Li superfluid at low $T$ with optical
molecular spectroscopy, and provide unambiguous evidence that $Z_{cc}$ is
governed by many-body physics.  We emphasize that a concrete relation
between $Z_{cc}$ and the density can be used to extract other physical
properties and to test various theories, and is thus much more
important than simply knowing $Z_{cc}\neq 0$ in the BCS regime.  Due to the
smallness of $Z_{cc}$~\cite{Partridge2005PRL}, precise control of
experimental parameters is needed in order to unravel the many-body
interaction effect.  Indeed, we find that due to the weak dependence
on particle number, the many-body effects could be easily buried in
noise, as was the case in Ref.~\cite{Partridge2005PRL}.  With advanced
laser cooling techniques, we are able to produce a $^6$Li superfluid with large atom number and very low temperature, which greatly improve the signal-to-noise ratio of the
measurements. Moreover, to reduce systematic errors, the Rabi frequency of the molecular transition is calibrated with the
well-known $Z_{cc}$ values in the BEC regime. With these improved techniques and
calibrations, in the unitary and BCS regimes, power-law fittings which account for
the many-body effects are in good accordance with the experimental
data, while obvious deviations are found for two-body-theory-based
exponential fittings, especially at high $B$.  At unitarity, the
universal relation $Z_{cc} =\eta \sqrt{T_\text{F}}$ has been revealed, with
$\eta= 0.074(12)$~K$^{-1/2}$, in quantitative agreement with the
theoretical prediction \cite{Chen2005PRL,ZhangSZ2009}.  (Note here K denotes Kelvin).  At a higher
field $B=925$~G in the near-BCS regime, we find a power-law exponent
much larger than 1/2, in quantitative agreement with
predictions as well. This higher exponent means a more sensitive
dependence of $Z_{cc}$ on $T_\text{F}$, and hence a stronger many-body
effect.  Furthermore, a proper treatment of thermal contributions of
the closed-channel molecules is of crucial importance. The data and
theory in the BCS regime can be brought into quantitative agreement by
taking into account of contributions of thermal noncondensed
closed-channel molecules outside the superfluid core in the trap.

\section{Results and discussion}
The experimental procedure for producing $^6$Li superfluid has been
described in our previous works~\cite{Yao2016PRL,Wu2017JPB}. The superfluid of $3.0 (1)\times 10^6$ $^6$Li atoms at $T/T_\text{F} = 0.05 (1)$ are confined in an oblate harmonic optical dipole trap (wavelength 1064~nm, 1/e$^2$ horizontal (vertical) radius 200~$\mu$m
(48~$\mu$m)). The radial confinement is mainly optical
with horizontal and vertical trap frequencies being
$2\pi\times$53.7(3)~Hz and $2\pi\times$205.3(5)~Hz, respectively. The
axial confinement is mainly provided by the magnetic field curvature
with a trap frequency of $2\pi\times$16.8(1) Hz at 832 G. The method for probing the closed-channel fraction $Z_{cc}$ is similar to
Ref.~\cite{Partridge2005PRL}, where a resonant laser transition is
used to pump the closed-channel molecules into an excited singlet
molecular state. Here, the transition is $X~^1\sum^+_g$ ($\nu=38$)
$\rightarrow$ $A~^1\sum^+_u$ ($\nu'=68$), since it possesses the largest
Franck-Condon wavefunction overlap. Due to Rabi oscillation and
spontaneous-emission loss, the number of dressed molecules decreases
at  rate $\Gamma=Z_{cc}\Omega_\text{m}^2/\gamma_\text{m}$, where $\Omega_\text{m}$ is the
Rabi frequency of the transition and $\gamma_\text{m}$ is the linewidth of
the excited molecular state. We mention that the $\Gamma$ expression is
valid provided $\Omega_\text{m}^2/\gamma_\text{m}^2 \ll 1$, which is easily satisfied
in our experiment, and has been further verified by varying the probe
laser intensity. Therefore, by recording the remaining atom number
versus probing time, the fraction $Z_{cc}$ can be extracted with given
$\Omega_\text{m}^2/\gamma_\text{m}$.

Previously, $\Omega_\text{m}^2/\gamma_\text{m}$ is directly calculated based on the
theoretical knowledge of molecular optical transition and the measured
laser beam parameters. This method relies on precise measurement of
laser power $P$ and beam waist $\omega_0$ at the position of the
atoms, which is very difficult to achieve in practice~\cite{Ravensbergen2018PRL}. Moreover, the calculated wave-function
overlap between the ground and excited molecular states is based
on some theoretical assumptions regarding molecular potentials. Thus,
the combination of these problems limits the accuracy of the obtained
$\Omega_\text{m}^2/\gamma_\text{m}$ to a few 10\%.  To solve this problem, the ratio
$\Omega_\text{m}^2/\gamma_\text{m}$ is determined by calibrating the measured $Z_{cc}$
against theory in the well understood deep BEC regime, where good
agreement between two-body and many-body theories have been achieved~\cite{Chen2005PRL}.  Indeed, in this regime, $Z_{cc}$ becomes independent
of $T_\text{F}$ \cite{Chen2005PRL}, so that the molecule number will decay
exponentially with a decay constant $\Gamma$ upon laser
exposure. Therefore, $\Omega_\text{m}^2/\gamma_\text{m}$ can be derived by linearly
fitting a series of theoretically calculated $Z_{cc}$ and experimentally
measured $\Gamma$ at different magnetic fields.

\begin{figure}
\centerline{\includegraphics[width=3.2in]{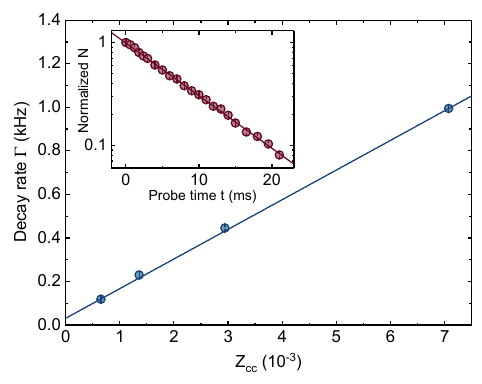}}
\caption{Calibration of the molecular transition parameter
  $\Omega_\text{m}^2/\gamma_\text{m}$. Shown is the decay rate $\Gamma$, measured at
  748, 722, 696, and 670~G, corresponding to the data points from left
  to right, respectively, versus theoretically calculated $Z_{cc}$.  Each
  data point is averaged over three measurements with error bars given
  by standard deviation. The slope of the linear fit yields
  $\Omega_\text{m}^2/\gamma_\text{m}$. The decay rate $\Gamma$ was obtained by
  exponential fitting of the remaining atoms $N(t)$ (normalized at
  $t=0$) as a function of the probe time $t$ (red line), as shown in
  the inset for 748~G.  }
\label{fig1}
\end{figure}

After preparing the $^6$Li superfluid at 832~G, the field is linearly
ramped to the desired value in 100~ms and held for another 100~ms for
equilibration before the molecular probing. The probe laser (laser
power 20~$\mu$W) is frequency-locked to an ultra-stable Fabry-Perot
cavity with its power stabilized. The achieved long-term stability of
frequency and power are 500~kHz and 0.1\%, respectively, which greatly
suppress the systematic errors. The measurements in the BEC regime are
performed at 670, 696, 722, and 748~G, respectively. Note that the
$1/e$ depletion time of milliseconds is far shorter than the molecule
lifetime of more than 10~s, thus the background molecule loss is
negligible. As an example, the inset of Fig.~\ref{fig1} shows the
remaining atom number as a function of probing time at 748~G, which
yields $\Gamma= 118.1(9)$~Hz. Plotted in Fig.~\ref{fig1} is the linear
fitting to $\Gamma$ versus theoretical $Z_{cc}$ values at these magnetic
fields, yielding a slope $\Omega_\text{m}^2/\gamma_\text{m}= 136(1)$~kHz. The
high-quality fitting curve demonstrates not only the stability of our
experimental setup, but also the reliability of obtained
$\Omega_\text{m}^2/\gamma_\text{m}$.

Before we move on to the measurements in the unitary and BCS regimes,
we summarize the theory predictions~\cite{Chen2005PRL} which
lay the foundation for our data analysis. The theory includes from the start the two-channel Feshbach physics, as described by the Hamiltonian in Ref.~\cite{JS2}.
At $T=0$ in a homogeneous Fermi gas, the superfluid order parameter
$\tilde{\Delta}$ associated with condensed dressed molecules has two
contributions, $\phi_\text{m}$ from the closed channel and $\Delta$
from the open channel, as $ \tilde{\Delta} = \Delta -g \phi_\text{m}$,
where $n_{b0} = \phi_\text{m}^2$ is the number of closed-channel
molecules, and $g$ is the inter-channel coupling~\cite{PhysRepReview}. Note that Cooper pairing in the BCS regime is
purely a many-body effect, and it is due to this linear combination
that the closed-channel molecules acquire a finite fraction in the BCS
regime. In the end, we have $n_{b0} = Z_g \tilde{\Delta}^2$, where the
coefficient $Z_g$ can be calculated using experimental parameters and
the fermionic chemical potential. Using a local density approximation,
the trap-averaged closed-channel fraction $Z_{cc}$, as
measured here, is thus given by $Z_{cc} = 2N_{b0}/N$ at low $T$, where
$N_{b0}$ and $N$ are trap integral of local $n_{b0}(r)$ and overall
atom density $n(r)$, respectively. The theory predicts that
$Z_{cc} =\eta \sqrt{T_\text{F}}$ at unitarity (832~G), with
$\eta = 0.066$~K$^{-1/2}$, where $1/k_\text{F}a=0$ holds for the whole
trap. Here $T_\text{F}=\hbar\overline{\omega}(3N)^{1/3}/k_B$, where
$N$ is the atom number and $\overline{\omega}$ is the geometric
average of trap frequencies.  Away from unitarity, the local
$1/k_\text{F}a$ become inhomogeneous across the trap, so that $Z_{cc}$ as a
function of $T_\text{F}$ can only be calculated numerically, as shown
in Fig.~3 of Ref.~\cite{Chen2005PRL}. Here we have recalculated $Z_{cc}$
using the most up-to-date resonance parameters~\cite{Zuern2013PRL}. A
log-log plot of $Z_{cc}$ versus $T_\text{F}$, presented in Supplementary
Fig.~S1, indicates that within the range of $T_\text{F}$ for our
experimental data, it can be reasonably approximated with a power-law
dependence, $Z_{cc}\propto T_\text{F}^\alpha$, with $\alpha>1/2$ in the BCS
regime.

For both the unitary and
BCS regimes, we shall write $Z_{cc}=\eta T_\text{F}^{\alpha} $. Substituting this into
the decay equation, one obtains
\begin{equation}
\dot{N}/N=-Z_{cc}\frac{\Omega_\text{m}^2}{\gamma_\text{m}}=-\beta N^{\alpha/3},
\label{Eq:differential}
\end{equation}
 where
$\beta=\eta ( 3^{1/3}\hbar \bar{\omega}/k_\text{B})^{\alpha} \Omega_\text{m}^2/\gamma_\text{m}$.
This leads to a power-law decay,
\begin{equation}\label{Eq:decay}
N(t)=\rho(t+\Delta t)^{-c} \,,
\end{equation}
where
$\alpha=3/c\,,\quad
\beta =c/\rho^{\alpha/3}$.
Then, the parameters $\eta$ and $\alpha$ can be acquired by fitting the
experimental data with Eq.~(\ref{Eq:decay}).

\begin{figure}
\centerline{\includegraphics[width=0.9\columnwidth]{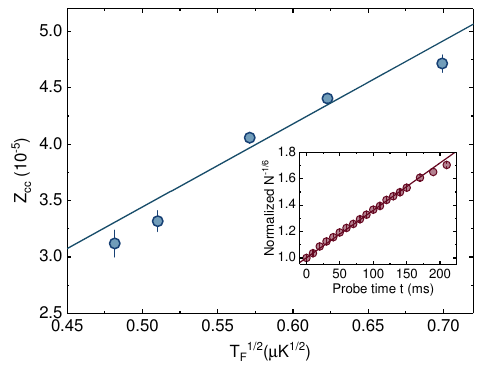}}
\caption{Measured $Z_{cc}$ as a function of $T_\text{F}^{1/2}$ at unitarity,
  exhibiting a good linearity.  Plotted in the inset is an example
  case of $N^{-1/6}$ versus $t$, where the remaining atoms $N$
  (normalized at $t=0$) are counted and statistically averaged over 3
  measurements. The vertical bars denote the standard error. Both the
  main figure and the inset yield nearly the same slope.}
\label{fig2}
\end{figure}

\begin{figure}
\centerline{\includegraphics[width=1.0\columnwidth]{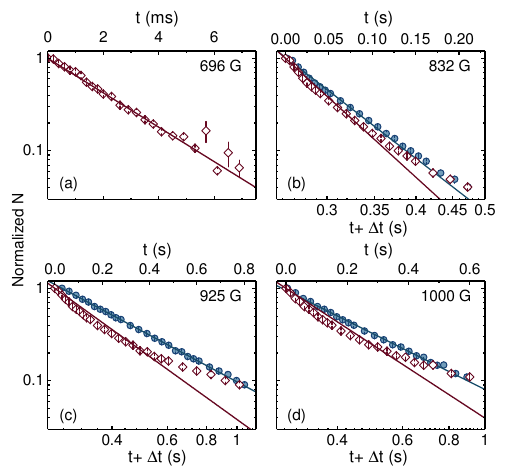}}
\caption{Optical molecular spectroscopy in the BCS-BEC crossover.
  Shown is the atom number $N$, normalized at $t=0$, measured at
  696, 832, 925 and 1000~G, as a function of $t$ on semi-log
  (red open diamonds, top axes) and log-log (blue filled circles,
  bottom axes) scales, with a laser power of (a) $20~\mu$W, (b)
  $60~\mu$W, (c) $120~\mu$W , and (d) $360~\mu$W, respectively. Error
  bars represent one sigma standard error. The straight lines are
  exponential (a-d, red) and (b-d, blue) power-law fits, respectively,
  which yields $\alpha = 1.68$ at 925~G and 2.10 at 1000~G.}
\label{fig3}
\end{figure}

Next, we measure $Z_{cc}$ in the unitary and BCS regimes with the
calibrated $\Omega_\text{m}^2/\gamma_\text{m}$. At unitarity,
$\alpha=1/2$, and thus $N(t)^{-1/6}\propto t+\Delta t$, with slope $\rho^{-1/6} =\beta /6 \propto \eta$. The $1/e$ decay time of
the molecules is carefully chosen to be about 60~ms, which is much
longer than the estimated equilibration time of the dressed molecules~\cite{Sanner2012PRL}. The inset of Fig.~\ref{fig2} shows as an example
$N^{-1/6}$ versus
$t$. The good agreement between the data and the linear fitting
demonstrates the validity of the theoretical model.  The fitted slope yields
$\eta=0.070 (1)$~K$^{-1/2}$, very close to the theory value
0.066~K$^{-1/2}$ at zero $T$.  For comparison, exponential fitting
clearly fails, as shown in Fig.~\ref{fig3}(b) below (open red diamonds),
despite that the change in $Z_{cc}$ is only about 30\% during the molecular
probing. We further perform a series of measurements of $Z_{cc}$ with
varying initial $T_\text{F}$, as plotted in Fig.~\ref{fig2}, which
exhibits a good proportionality between $Z_{cc}$ and $\sqrt{T_\text{F}}$, with a
coefficient of 0.074(12)~K$^{-1/2}$, in quantitative agreement with that
from a single set of probing data in the inset. In contrast, the
measurements and analysis in Ref.~\cite{Partridge2005PRL} allows only
one value of $Z_{cc}$ for each interaction strength.

With the same procedure, we probe $Z_{cc}$ on the BCS side for
$B=850\sim 1000$~G.  In these measurements, to eliminate the potential non-equilibrium effects caused by magnetic field ramping, the Fermi gas is evaporatively cooled at the same field $B$ for molecular probing. At unitarity, the system temperature can be determined by fitting the \textit{in situ} density distribution with the known equation of state (EoS)~\cite{Zwierlein2011}. Unfortunately, we cannot quantitatively determine the temperature at higher magnetic fields, due to the lack of reliable knowledge of the EoS in the crossover region. Nevertheless, a roughly linear increase in temperature with the magnetic field could be inferred from the observed cloud size change through time-of-flight measurement~\cite{Ko2019NP}. We attribute the slight increase of temperature to the decrease of elastic scattering rate in the BCS regime. The rapid decrease of $Z_{cc}$ with
$B$ leads to a significant increase of the $1/e$ depletion time. To
suppress the influence of background loss, we increase the laser power
gradually from 60~$\mu$W to 360~$\mu$W to maintain an approximately
identical ``decay constant'' of about 200~ms throughout the whole BCS
regime. In Fig.~\ref{fig3}, we plot $N$ (normalized at $t=0$) as a
function of $t$ for $B= 696$, 832, 925, and 1000~G, which spans from
the BEC, the unitary, to the BCS regimes. While the exponential decay
function fits well with the data at 696~G in the BEC regime
(Fig.~\ref{fig3}(a)), there is a progressively increasing systematic
deviation as the field increases (red open diamonds). The failure of
the exponential fitting (semi-log scales, top axes) becomes obvious in
the unitary and BCS regimes.  In contrast, the power-law fitting with
increasing exponent works perfectly well for the unitary and BCS
cases, as manifested by the good straight fitting lines in log-log
scales (blue solid circles, bottom axes) in
Fig.~\ref{fig3}(b)-(d). These results provide direct evidence of the
many-body effects in $Z_{cc}$.

\begin{figure}
\centerline{\includegraphics[width=0.95\columnwidth]{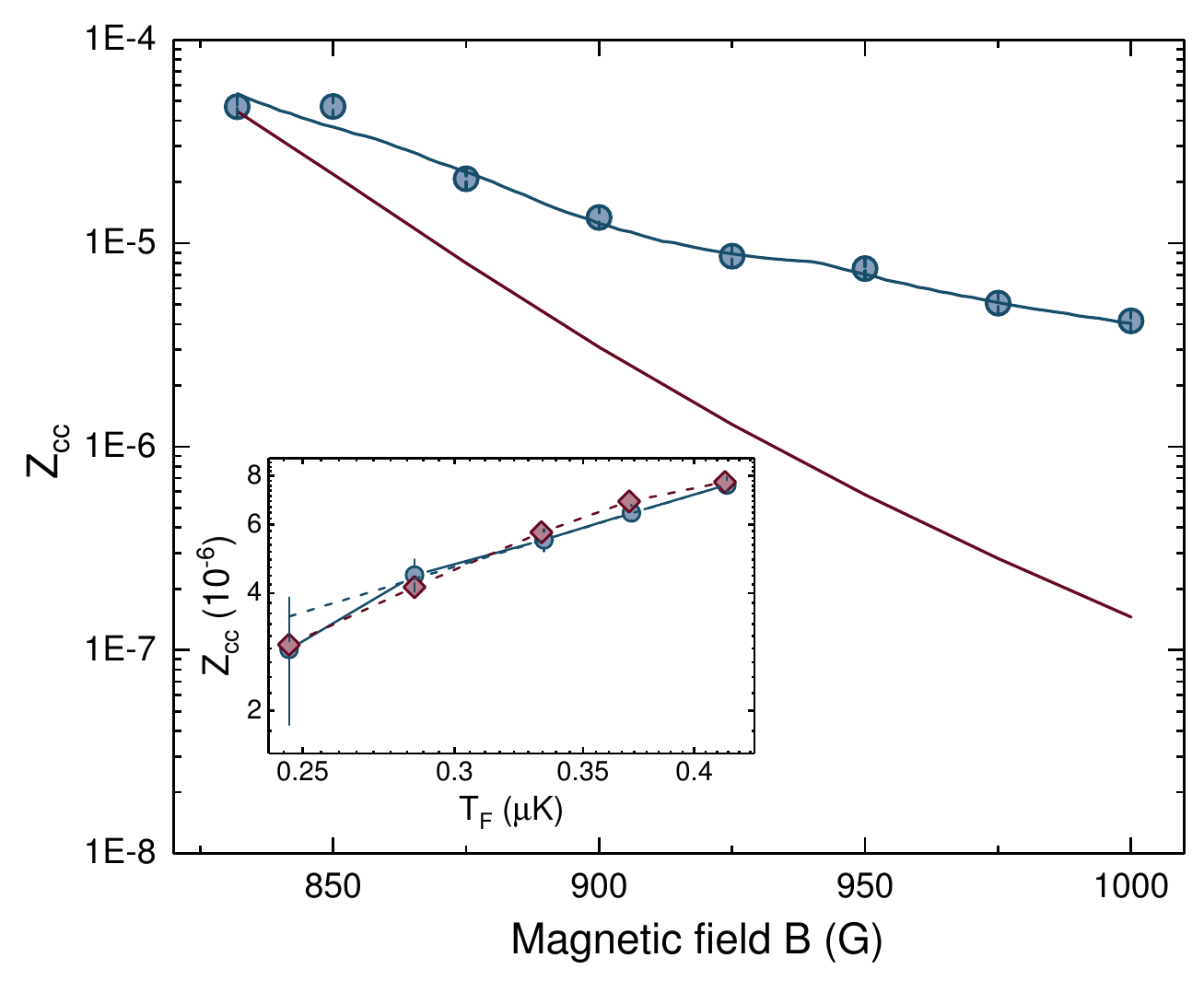}}
\caption{Measured closed-channel fraction $Z_{cc}$ as a function of
  magnetic field for $T_\text{F}=0.45~\mu$K between experiment (blue
  solid circles) and theory at $T=0$ (red curve) and at linearly
  varying $T/T_\text{F}$ from 0.034 at 832~G to 0.056 at 1000~G (blue
  curve).  Shown in the inset is $Z_{cc}$ (blue solid circles) versus $T_\text{F}$ at 925~G at
  low $T$ in log-log scale. The blue dashed line is a power law fit
  with exponent $\alpha = 1.48(8)$.  For comparison, also plotted are
  the theoretical values (red diamonds) calculated at $T/T_\text{F}$
  linearly varying between $0.046$ at $T_\text{F}=0.416$~$\mu$K and
  0.049 at $T_\text{F}=0.246$~$\mu$K.  }
\label{fig4}
\end{figure}

In Fig.~\ref{fig4}, we compare the experimental and
theoretical $Z_{cc}$ values for $T_\text{F}=0.45~\mu$K on the BCS side of the
FR. In the vicinity of unitarity, our experimental results are in good agreement with the many-body theoretical
predictions calculated at $T=0$. However, as the field increases, a progressively increasing departure between theory and experiment is found (see red curve in Fig.~\ref{fig4}), which has also been noticed in Ref.~\cite{Chen2005PRL}. This is because in the BCS regime, the dispersion of finite momentum closed channel molecules becomes significantly softened such that the population of non-condensed molecules at the trap edge becomes dramatically enhanced at finite $T$. Thus, it is crucial to treat properly the thermal contributions of the closed-channel molecules, especially outside the superfluid core, in the BCS regime (see Supplementary Information for theoretical details). With theoretical improvements, semi-quantitative agreement between theory and experiment has been achieved, by assuming a finite but reasonable temperature (blue solid curve, $T/T_\text{F}$ linearly varying from $0.034$ at 832~G to 0.056 at 1000~G).


Presented in the inset of Fig.~\ref{fig4} is the measured $Z_{cc}$ versus
$T_\text{F}$ at 925~G at low $T$ in log-log scale.
The power law fit (blue dashed line) yields an exponent
$\alpha = 1.48(8)$, which is consistent with that obtained from
Fig.~\ref{fig3}(c), but much larger than $1/2$ at unitarity. (There
seems to be a slight curvature that agrees with the
theory curve in Fig.~S1). For comparison, also plotted are the
theoretically calculated values (blue squares) with $T/T_\text{F}$
linearly varying between $0.046$ at $T_\text{F}=0.416$~$\mu$K and
0.049 at $T_\text{F}=0.246$~$\mu$K, which exhibits a
(semi-)quantitative agreement with experiment.  This slight
temperature variation of $T/T_\text{F}$ is reasonable since
$T/T_\text{F}$ was in fact slightly higher at lower $T_\text{F}$.

Finally, we point out that, in a two-channel model, the density (or
equivalently $T_\text{F}$) provides an extra dimension to the system. More
specifically, two Fermi gases with different $T_\text{F}$ are no longer
mathematically equivalent, even if they share the same
$1/k_\text{F}a$ and
$T/T_\text{F}$. This will
inevitably lead to violation, albeit small, of the universality
hypothesis of a unitary Fermi gas
\cite{Heiselberg2001PRA,Carlson2003PRL,Ho2004PRL}, which is based on a
one-channel assumption.  At the same time, over the entire BCS-BEC
crossover regimes, it is predicted that $Z_{cc}\propto \Delta^2$ roughly
for given $T_\text{F}$ and interaction strength
\cite{Chen2005PRL}. Furthermore, $Z_{cc}$ also exhibits important
$T$ dependence \cite{Chen2005PRL}. Hence measurement of
condensed and noncondensed closed-channel fractions as a function of
$T$ may disclose how the order parameter and the pseudogap evolve with
temperature.

Note that we do not extract the (trap averaged) Tan's contact
\cite{Tan2008} since its precise value is not yet available for
comparison as a function of $B$ and $T$, especially in a trap.

\section{Conclusion}
In summary, we have measured $Z_{cc}$ of interacting Fermi gases of $^6$Li
in a trap as a function of $T_\text{F}$ and $B$. Away from the deep
BEC regime, the fraction $Z_{cc}$ exhibits clear dependence on Fermi
temperature $T_\text{F}$, unraveling important many-body interaction
effect.  In particular, we obtain $Z_{cc} =\eta \sqrt{T_\text{F}}$ at
unitarity, with  $\eta= 0.074(12)$~K$^{-1/2}$, in
quantitative agreement with theory. It would be interesting to perform
precision test of the universality hypothesis, investigate the
temperature evolution of the condensed and noncondensed part of the
closed-channel fraction, and test against different BCS-BEC crossover
theories in the future.

\section{Acknowledgments}
We thank R. Hulet for valuable communications.

\section{Funding}
This work is supported by the National Key R\&D Program of China (Grant No. 2018YFA0306501), National Natural Science Foundation of China (Grant Nos. 11874340, 11425417, 11774309), the Chinese Academy of Sciences (CAS), the Anhui Initiative in Quantum Information Technologies, the Shanghai Municipal Science and Technology Major Project (Grant No. 2019SHZDZX01), Natural Science Foundation of Zhejiang Province of China (Grant No.
LZ13A040001). K.L. was supported by the National Science Foundation (Grant No. NSF-DMR-MRSEC 1420709).
\bibliography{closedchannel}

\begin{thebibliography}{37}%
\makeatletter
\providecommand \@ifxundefined [1]{%
 \@ifx{#1\undefined}
}%
\providecommand \@ifnum [1]{%
 \ifnum #1\expandafter \@firstoftwo
 \else \expandafter \@secondoftwo
 \fi
}%
\providecommand \@ifx [1]{%
 \ifx #1\expandafter \@firstoftwo
 \else \expandafter \@secondoftwo
 \fi
}%
\providecommand \natexlab [1]{#1}%
\providecommand \enquote  [1]{``#1''}%
\providecommand \bibnamefont  [1]{#1}%
\providecommand \bibfnamefont [1]{#1}%
\providecommand \citenamefont [1]{#1}%
\providecommand \href@noop [0]{\@secondoftwo}%
\providecommand \href [0]{\begingroup \@sanitize@url \@href}%
\providecommand \@href[1]{\@@startlink{#1}\@@href}%
\providecommand \@@href[1]{\endgroup#1\@@endlink}%
\providecommand \@sanitize@url [0]{\catcode `\\12\catcode `\$12\catcode
  `\&12\catcode `\#12\catcode `\^12\catcode `\_12\catcode `\%12\relax}%
\providecommand \@@startlink[1]{}%
\providecommand \@@endlink[0]{}%
\providecommand \url  [0]{\begingroup\@sanitize@url \@url }%
\providecommand \@url [1]{\endgroup\@href {#1}{\urlprefix }}%
\providecommand \urlprefix  [0]{URL }%
\providecommand \Eprint [0]{\href }%
\providecommand \doibase [0]{http://dx.doi.org/}%
\providecommand \selectlanguage [0]{\@gobble}%
\providecommand \bibinfo  [0]{\@secondoftwo}%
\providecommand \bibfield  [0]{\@secondoftwo}%
\providecommand \translation [1]{[#1]}%
\providecommand \BibitemOpen [0]{}%
\providecommand \bibitemStop [0]{}%
\providecommand \bibitemNoStop [0]{.\EOS\space}%
\providecommand \EOS [0]{\spacefactor3000\relax}%
\providecommand \BibitemShut  [1]{\csname bibitem#1\endcsname}%
\let\auto@bib@innerbib\@empty
\bibitem [{\citenamefont {Bloch}\ \emph {et~al.}(2008)\citenamefont {Bloch},
  \citenamefont {Dalibard},\ and\ \citenamefont {Zwerger}}]{Bloch2008RoMP}%
  \BibitemOpen
  \bibfield  {author} {\bibinfo {author} {\bibfnamefont {I.}~\bibnamefont
  {Bloch}}, \bibinfo {author} {\bibfnamefont {J.}~\bibnamefont {Dalibard}}, \
  and\ \bibinfo {author} {\bibfnamefont {W.}~\bibnamefont {Zwerger}},\
  }\bibfield  {title} {\enquote {\bibinfo {title} {Many-body physics with
  ultracold gases},}\ }\href {\doibase 10.1103/revmodphys.80.885} {\bibfield
  {journal} {\bibinfo  {journal} {Rev. Mod. Phys.}\ }\textbf {\bibinfo {volume}
  {80}},\ \bibinfo {pages} {885--964} (\bibinfo {year} {2008})}\BibitemShut
  {NoStop}%
\bibitem [{\citenamefont {Leggett}(1980)}]{Leggett}%
  \BibitemOpen
  \bibfield  {author} {\bibinfo {author} {\bibfnamefont {A.~J.}\ \bibnamefont
  {Leggett}},\ }\bibfield  {title} {\enquote {\bibinfo {title} {Diatomic
  molecules and {Cooper} pairs},}\ }in\ \href
  {https://link.springer.com/chapter/10.1007/BFb0120125} {\emph {\bibinfo
  {booktitle} {Modern Trends in the Theory of Condensed Matter}}}\ (\bibinfo
  {publisher} {Springer-Verlag},\ \bibinfo {address} {Berlin},\ \bibinfo {year}
  {1980})\ pp.\ \bibinfo {pages} {13--27}\BibitemShut {NoStop}%
\bibitem [{\citenamefont {Chen}\ \emph {et~al.}(2005)\citenamefont {Chen},
  \citenamefont {Stajic}, \citenamefont {Tan},\ and\ \citenamefont
  {Levin}}]{PhysRepReview}%
  \BibitemOpen
  \bibfield  {author} {\bibinfo {author} {\bibfnamefont {Q.}~\bibnamefont
  {Chen}}, \bibinfo {author} {\bibfnamefont {J.}~\bibnamefont {Stajic}},
  \bibinfo {author} {\bibfnamefont {S.}~\bibnamefont {Tan}}, \ and\ \bibinfo
  {author} {\bibfnamefont {K.}~\bibnamefont {Levin}},\ }\bibfield  {title}
  {\enquote {\bibinfo {title} {{BCS--BEC} crossover: From high temperature
  superconductors to ultracold superfluids},}\ }\href {\doibase
  10.1016/j.physrep.2005.02.005} {\bibfield  {journal} {\bibinfo  {journal}
  {Phys. Rep.}\ }\textbf {\bibinfo {volume} {412}},\ \bibinfo {pages} {1--88}
  (\bibinfo {year} {2005})}\BibitemShut {NoStop}%
\bibitem [{\citenamefont {Jochim}\ \emph {et~al.}(2003)\citenamefont {Jochim},
  \citenamefont {Bartenstein}, \citenamefont {Altmeyer}, \citenamefont {Hendl},
  \citenamefont {Riedl}, \citenamefont {Chin}, \citenamefont {Denschlag},\ and\
  \citenamefont {Grimm}}]{Jochim2003S}%
  \BibitemOpen
  \bibfield  {author} {\bibinfo {author} {\bibfnamefont {S.}~\bibnamefont
  {Jochim}}, \bibinfo {author} {\bibfnamefont {M.}~\bibnamefont {Bartenstein}},
  \bibinfo {author} {\bibfnamefont {A.}~\bibnamefont {Altmeyer}}, \bibinfo
  {author} {\bibfnamefont {G.}~\bibnamefont {Hendl}}, \bibinfo {author}
  {\bibfnamefont {S.}~\bibnamefont {Riedl}}, \bibinfo {author} {\bibfnamefont
  {C.}~\bibnamefont {Chin}}, \bibinfo {author} {\bibfnamefont {J.~H.}\
  \bibnamefont {Denschlag}}, \ and\ \bibinfo {author} {\bibfnamefont
  {R.}~\bibnamefont {Grimm}},\ }\bibfield  {title} {\enquote {\bibinfo {title}
  {{Bose-Einstein} condensation of molecules},}\ }\href {\doibase
  10.1126/science.1093280} {\bibfield  {journal} {\bibinfo  {journal}
  {Science}\ }\textbf {\bibinfo {volume} {302}},\ \bibinfo {pages} {2101--2103}
  (\bibinfo {year} {2003})}\BibitemShut {NoStop}%
\bibitem [{\citenamefont {Greiner}\ \emph {et~al.}(2003)\citenamefont
  {Greiner}, \citenamefont {Regal},\ and\ \citenamefont {Jin}}]{Greiner2003N}%
  \BibitemOpen
  \bibfield  {author} {\bibinfo {author} {\bibfnamefont {M.}~\bibnamefont
  {Greiner}}, \bibinfo {author} {\bibfnamefont {C.~A.}\ \bibnamefont {Regal}},
  \ and\ \bibinfo {author} {\bibfnamefont {D.~S.}\ \bibnamefont {Jin}},\
  }\bibfield  {title} {\enquote {\bibinfo {title} {Emergence of a molecular
  {Bose-Einstein} condensate from a {F}ermi gas},}\ }\href {\doibase
  10.1038/nature02199} {\bibfield  {journal} {\bibinfo  {journal} {Nature}\
  }\textbf {\bibinfo {volume} {426}},\ \bibinfo {pages} {537--540} (\bibinfo
  {year} {2003})}\BibitemShut {NoStop}%
\bibitem [{\citenamefont {Kinast}\ \emph {et~al.}(2005)\citenamefont {Kinast},
  \citenamefont {Turlapow}, \citenamefont {Thomas}, \citenamefont {Chen},
  \citenamefont {Stajic},\ and\ \citenamefont {Levin}}]{Kinast2005S}%
  \BibitemOpen
  \bibfield  {author} {\bibinfo {author} {\bibfnamefont {J.}~\bibnamefont
  {Kinast}}, \bibinfo {author} {\bibfnamefont {A.}~\bibnamefont {Turlapow}},
  \bibinfo {author} {\bibfnamefont {J.~E.}\ \bibnamefont {Thomas}}, \bibinfo
  {author} {\bibfnamefont {Q.~J.}\ \bibnamefont {Chen}}, \bibinfo {author}
  {\bibfnamefont {J.}~\bibnamefont {Stajic}}, \ and\ \bibinfo {author}
  {\bibfnamefont {K.}~\bibnamefont {Levin}},\ }\bibfield  {title} {\enquote
  {\bibinfo {title} {Heat capacity of a strongly interacting {F}ermi gas},}\
  }\href {\doibase 10.1126/science.1109220} {\bibfield  {journal} {\bibinfo
  {journal} {Science}\ }\textbf {\bibinfo {volume} {307}},\ \bibinfo {pages}
  {1296--1299} (\bibinfo {year} {2005})}\BibitemShut {NoStop}%
\bibitem [{\citenamefont {Navon}\ \emph {et~al.}(2010)\citenamefont {Navon},
  \citenamefont {Nascimbene}, \citenamefont {Chevy},\ and\ \citenamefont
  {Salomon}}]{Navon2010S}%
  \BibitemOpen
  \bibfield  {author} {\bibinfo {author} {\bibfnamefont {N.}~\bibnamefont
  {Navon}}, \bibinfo {author} {\bibfnamefont {S.}~\bibnamefont {Nascimbene}},
  \bibinfo {author} {\bibfnamefont {F.}~\bibnamefont {Chevy}}, \ and\ \bibinfo
  {author} {\bibfnamefont {C.}~\bibnamefont {Salomon}},\ }\bibfield  {title}
  {\enquote {\bibinfo {title} {The equation of state of a low-temperature
  {F}ermi gas with tunable interactions},}\ }\href {\doibase
  10.1126/science.1187582} {\bibfield  {journal} {\bibinfo  {journal}
  {Science}\ }\textbf {\bibinfo {volume} {328}},\ \bibinfo {pages} {729--732}
  (\bibinfo {year} {2010})}\BibitemShut {NoStop}%
\bibitem [{\citenamefont {Nascimb{\`{e}}ne}\ \emph {et~al.}(2010)\citenamefont
  {Nascimb{\`{e}}ne}, \citenamefont {Navon}, \citenamefont {Jiang},
  \citenamefont {Chevy},\ and\ \citenamefont {Salomon}}]{Nascimbene2010N}%
  \BibitemOpen
  \bibfield  {author} {\bibinfo {author} {\bibfnamefont {S.}~\bibnamefont
  {Nascimb{\`{e}}ne}}, \bibinfo {author} {\bibfnamefont {N.}~\bibnamefont
  {Navon}}, \bibinfo {author} {\bibfnamefont {K.~J.}\ \bibnamefont {Jiang}},
  \bibinfo {author} {\bibfnamefont {F.}~\bibnamefont {Chevy}}, \ and\ \bibinfo
  {author} {\bibfnamefont {C.}~\bibnamefont {Salomon}},\ }\bibfield  {title}
  {\enquote {\bibinfo {title} {Exploring the thermodynamics of a universal
  {F}ermi gas},}\ }\href {\doibase 10.1038/nature08814} {\bibfield  {journal}
  {\bibinfo  {journal} {Nature}\ }\textbf {\bibinfo {volume} {463}},\ \bibinfo
  {pages} {1057--1060} (\bibinfo {year} {2010})}\BibitemShut {NoStop}%
\bibitem [{\citenamefont {Ku}\ \emph {et~al.}(2014)\citenamefont {Ku},
  \citenamefont {Ji}, \citenamefont {Mukherjee}, \citenamefont
  {Guardado-Sanchez}, \citenamefont {Cheuk}, \citenamefont {Yefsah},\ and\
  \citenamefont {Zwierlein}}]{Ku2014PRL}%
  \BibitemOpen
  \bibfield  {author} {\bibinfo {author} {\bibfnamefont
  {M.~J.{\hspace{0.167em}}H.}\ \bibnamefont {Ku}}, \bibinfo {author}
  {\bibfnamefont {W.}~\bibnamefont {Ji}}, \bibinfo {author} {\bibfnamefont
  {B.}~\bibnamefont {Mukherjee}}, \bibinfo {author} {\bibfnamefont
  {E.}~\bibnamefont {Guardado-Sanchez}}, \bibinfo {author} {\bibfnamefont
  {L.~W.}\ \bibnamefont {Cheuk}}, \bibinfo {author} {\bibfnamefont
  {T.}~\bibnamefont {Yefsah}}, \ and\ \bibinfo {author} {\bibfnamefont {M.~W.}\
  \bibnamefont {Zwierlein}},\ }\bibfield  {title} {\enquote {\bibinfo {title}
  {Motion of a solitonic vortex in the {BEC}-{BCS} crossover},}\ }\href
  {\doibase 10.1103/physrevlett.113.065301} {\bibfield  {journal} {\bibinfo
  {journal} {Phys. Rev. Lett.}\ }\textbf {\bibinfo {volume} {113}},\ \bibinfo
  {pages} {065301} (\bibinfo {year} {2014})}\BibitemShut {NoStop}%
\bibitem [{\citenamefont {Ries}\ \emph {et~al.}(2015)\citenamefont {Ries},
  \citenamefont {Wenz}, \citenamefont {Z\"{u}rn}, \citenamefont {Bayha},
  \citenamefont {Boettcher}, \citenamefont {Kedar}, \citenamefont {Murthy},
  \citenamefont {Neidig}, \citenamefont {Lompe},\ and\ \citenamefont
  {Jochim}}]{Ries2015PRL}%
  \BibitemOpen
  \bibfield  {author} {\bibinfo {author} {\bibfnamefont {M.~G.}\ \bibnamefont
  {Ries}}, \bibinfo {author} {\bibfnamefont {A.~N.}\ \bibnamefont {Wenz}},
  \bibinfo {author} {\bibfnamefont {G.}~\bibnamefont {Z\"{u}rn}}, \bibinfo
  {author} {\bibfnamefont {L.}~\bibnamefont {Bayha}}, \bibinfo {author}
  {\bibfnamefont {I.}~\bibnamefont {Boettcher}}, \bibinfo {author}
  {\bibfnamefont {D.}~\bibnamefont {Kedar}}, \bibinfo {author} {\bibfnamefont
  {P.~A.}\ \bibnamefont {Murthy}}, \bibinfo {author} {\bibfnamefont
  {M.}~\bibnamefont {Neidig}}, \bibinfo {author} {\bibfnamefont
  {T.}~\bibnamefont {Lompe}}, \ and\ \bibinfo {author} {\bibfnamefont
  {S.}~\bibnamefont {Jochim}},\ }\bibfield  {title} {\enquote {\bibinfo {title}
  {Observation of pair condensation in the quasi-2d {BEC}-{BCS} crossover},}\
  }\href {\doibase 10.1103/physrevlett.114.230401} {\bibfield  {journal}
  {\bibinfo  {journal} {Phys. Rev. Lett.}\ }\textbf {\bibinfo {volume} {114}},\
  \bibinfo {pages} {230401} (\bibinfo {year} {2015})}\BibitemShut {NoStop}%
\bibitem [{\citenamefont {Boettcher}\ \emph {et~al.}(2016)\citenamefont
  {Boettcher}, \citenamefont {Bayha}, \citenamefont {Kedar}, \citenamefont
  {Murthy}, \citenamefont {Neidig}, \citenamefont {Ries}, \citenamefont {Wenz},
  \citenamefont {Z\"{u}rn}, \citenamefont {Jochim},\ and\ \citenamefont
  {Enss}}]{Boettcher2016PRL}%
  \BibitemOpen
  \bibfield  {author} {\bibinfo {author} {\bibfnamefont {I.}~\bibnamefont
  {Boettcher}}, \bibinfo {author} {\bibfnamefont {L.}~\bibnamefont {Bayha}},
  \bibinfo {author} {\bibfnamefont {D.}~\bibnamefont {Kedar}}, \bibinfo
  {author} {\bibfnamefont {P.~A.}\ \bibnamefont {Murthy}}, \bibinfo {author}
  {\bibfnamefont {M.}~\bibnamefont {Neidig}}, \bibinfo {author} {\bibfnamefont
  {M.~G.}\ \bibnamefont {Ries}}, \bibinfo {author} {\bibfnamefont {A.~N.}\
  \bibnamefont {Wenz}}, \bibinfo {author} {\bibfnamefont {G.}~\bibnamefont
  {Z\"{u}rn}}, \bibinfo {author} {\bibfnamefont {S.}~\bibnamefont {Jochim}}, \
  and\ \bibinfo {author} {\bibfnamefont {T.}~\bibnamefont {Enss}},\ }\bibfield
  {title} {\enquote {\bibinfo {title} {Equation of state of ultracold fermions
  in the 2d {BEC}-{BCS} crossover region},}\ }\href {\doibase
  10.1103/physrevlett.116.045303} {\bibfield  {journal} {\bibinfo  {journal}
  {Phys. Rev. Lett.}\ }\textbf {\bibinfo {volume} {116}},\ \bibinfo {pages}
  {045303} (\bibinfo {year} {2016})}\BibitemShut {NoStop}%
\bibitem [{\citenamefont {Duine}\ and\ \citenamefont
  {Stoof}(2003)}]{Duine2003PRL}%
  \BibitemOpen
  \bibfield  {author} {\bibinfo {author} {\bibfnamefont {R.~A.}\ \bibnamefont
  {Duine}}\ and\ \bibinfo {author} {\bibfnamefont {H.~T.~C.}\ \bibnamefont
  {Stoof}},\ }\bibfield  {title} {\enquote {\bibinfo {title} {Many-body aspects
  of coherent atom-molecule oscillations},}\ }\href {\doibase
  10.1103/physrevlett.91.150405} {\bibfield  {journal} {\bibinfo  {journal}
  {Phys. Rev. Lett.}\ }\textbf {\bibinfo {volume} {91}},\ \bibinfo {pages}
  {150405} (\bibinfo {year} {2003})}\BibitemShut {NoStop}%
\bibitem [{\citenamefont {Duine}\ and\ \citenamefont
  {Stoof}(2004)}]{Duine2004PR}%
  \BibitemOpen
  \bibfield  {author} {\bibinfo {author} {\bibfnamefont {R.A.}\ \bibnamefont
  {Duine}}\ and\ \bibinfo {author} {\bibfnamefont {H.T.C.}\ \bibnamefont
  {Stoof}},\ }\bibfield  {title} {\enquote {\bibinfo {title}
  {Atom{\textendash}molecule coherence in bose gases},}\ }\href {\doibase
  10.1016/j.physrep.2004.03.003} {\bibfield  {journal} {\bibinfo  {journal}
  {Phys. Rep.}\ }\textbf {\bibinfo {volume} {396}},\ \bibinfo {pages}
  {115--195} (\bibinfo {year} {2004})}\BibitemShut {NoStop}%
\bibitem [{\citenamefont {Romans}\ and\ \citenamefont
  {Stoof}(2005)}]{Romans2005PRL}%
  \BibitemOpen
  \bibfield  {author} {\bibinfo {author} {\bibfnamefont {M.~W.~J.}\
  \bibnamefont {Romans}}\ and\ \bibinfo {author} {\bibfnamefont {H.~T.~C.}\
  \bibnamefont {Stoof}},\ }\bibfield  {title} {\enquote {\bibinfo {title}
  {Dressed {Feshbach} molecules in the {BEC}--{BCS} crossover},}\ }\href
  {\doibase 10.1103/physrevlett.95.260407} {\bibfield  {journal} {\bibinfo
  {journal} {Phys. Rev. Lett.}\ }\textbf {\bibinfo {volume} {95}},\ \bibinfo
  {pages} {260407} (\bibinfo {year} {2005})}\BibitemShut {NoStop}%
\bibitem [{\citenamefont {Chin}\ \emph {et~al.}(2004)\citenamefont {Chin},
  \citenamefont {Bartenstein}, \citenamefont {Altmeyer}, \citenamefont {Riedl},
  \citenamefont {Jochim}, \citenamefont {Denschlag},\ and\ \citenamefont
  {Grimm}}]{Grimm2004S}%
  \BibitemOpen
  \bibfield  {author} {\bibinfo {author} {\bibfnamefont {C.}~\bibnamefont
  {Chin}}, \bibinfo {author} {\bibfnamefont {M.}~\bibnamefont {Bartenstein}},
  \bibinfo {author} {\bibfnamefont {A.}~\bibnamefont {Altmeyer}}, \bibinfo
  {author} {\bibfnamefont {S.}~\bibnamefont {Riedl}}, \bibinfo {author}
  {\bibfnamefont {S.}~\bibnamefont {Jochim}}, \bibinfo {author} {\bibfnamefont
  {J.~H.}\ \bibnamefont {Denschlag}}, \ and\ \bibinfo {author} {\bibfnamefont
  {R.}~\bibnamefont {Grimm}},\ }\bibfield  {title} {\enquote {\bibinfo {title}
  {Observation of the pairing gap in a strongly interacting {F}ermi gas
  superfluids},}\ }\href {\doibase 10.1126/science.1100818} {\bibfield
  {journal} {\bibinfo  {journal} {Science}\ }\textbf {\bibinfo {volume}
  {305}},\ \bibinfo {pages} {1128--1130} (\bibinfo {year} {2004})}\BibitemShut
  {NoStop}%
\bibitem [{\citenamefont {Chen}\ and\ \citenamefont
  {Levin}(2005)}]{Chen2005PRL}%
  \BibitemOpen
  \bibfield  {author} {\bibinfo {author} {\bibfnamefont {Q.}~\bibnamefont
  {Chen}}\ and\ \bibinfo {author} {\bibfnamefont {K.}~\bibnamefont {Levin}},\
  }\bibfield  {title} {\enquote {\bibinfo {title} {Population of closed-channel
  molecules in trapped {F}ermi gases with broad {F}eshbach resonances},}\
  }\href {\doibase 10.1103/physrevlett.95.260406} {\bibfield  {journal}
  {\bibinfo  {journal} {Phys. Rev. Lett.}\ }\textbf {\bibinfo {volume} {95}},\
  \bibinfo {pages} {260406} (\bibinfo {year} {2005})}\BibitemShut {NoStop}%
\bibitem [{\citenamefont {Stewart}\ \emph {et~al.}(2008)\citenamefont
  {Stewart}, \citenamefont {Gaebler},\ and\ \citenamefont {Jin}}]{Jin2008N}%
  \BibitemOpen
  \bibfield  {author} {\bibinfo {author} {\bibfnamefont {J.~T.}\ \bibnamefont
  {Stewart}}, \bibinfo {author} {\bibfnamefont {J.~P.}\ \bibnamefont
  {Gaebler}}, \ and\ \bibinfo {author} {\bibfnamefont {D.~S.}\ \bibnamefont
  {Jin}},\ }\bibfield  {title} {\enquote {\bibinfo {title} {Using photoemission
  spectroscopy to probe a strongly interacting {Fermi} gas},}\ }\href {\doibase
  10.1038/nature07172} {\bibfield  {journal} {\bibinfo  {journal} {Nature
  (London)}\ }\textbf {\bibinfo {volume} {454}},\ \bibinfo {pages} {744}
  (\bibinfo {year} {2008})}\BibitemShut {NoStop}%
\bibitem [{\citenamefont {Tan}(2008)}]{Tan2008}%
  \BibitemOpen
  \bibfield  {author} {\bibinfo {author} {\bibfnamefont {S.}~\bibnamefont
  {Tan}},\ }\bibfield  {title} {\enquote {\bibinfo {title} {{Large momentum
  part of a strongly correlated Fermi gas}},}\ }\href {\doibase
  10.1016/j.aop.2008.03.005} {\bibfield  {journal} {\bibinfo  {journal} {Ann.
  Phys.}\ }\textbf {\bibinfo {volume} {323}},\ \bibinfo {pages} {2971--2986}
  (\bibinfo {year} {2008})}\BibitemShut {NoStop}%
\bibitem [{\citenamefont {Werner}\ \emph {et~al.}(2009)\citenamefont {Werner},
  \citenamefont {Tarruell},\ and\ \citenamefont {Castin}}]{Werner2009TEPJB}%
  \BibitemOpen
  \bibfield  {author} {\bibinfo {author} {\bibfnamefont {F.}~\bibnamefont
  {Werner}}, \bibinfo {author} {\bibfnamefont {L.}~\bibnamefont {Tarruell}}, \
  and\ \bibinfo {author} {\bibfnamefont {Y.}~\bibnamefont {Castin}},\
  }\bibfield  {title} {\enquote {\bibinfo {title} {Number of closed-channel
  molecules in the {BEC}--{BCS} crossover},}\ }\href {\doibase
  10.1140/epjb/e2009-00040-8} {\bibfield  {journal} {\bibinfo  {journal} {Eur.
  Phys. J. B}\ }\textbf {\bibinfo {volume} {68}},\ \bibinfo {pages} {401--415}
  (\bibinfo {year} {2009})}\BibitemShut {NoStop}%
\bibitem [{\citenamefont {Sagi}\ \emph {et~al.}(2012)\citenamefont {Sagi},
  \citenamefont {Drake}, \citenamefont {Paudel},\ and\ \citenamefont
  {Jin}}]{Sagi2012PRL}%
  \BibitemOpen
  \bibfield  {author} {\bibinfo {author} {\bibfnamefont {Y.}~\bibnamefont
  {Sagi}}, \bibinfo {author} {\bibfnamefont {T.~E.}\ \bibnamefont {Drake}},
  \bibinfo {author} {\bibfnamefont {R.}~\bibnamefont {Paudel}}, \ and\ \bibinfo
  {author} {\bibfnamefont {D.~S.}\ \bibnamefont {Jin}},\ }\bibfield  {title}
  {\enquote {\bibinfo {title} {Measurement of the homogeneous contact of a
  unitary {Fermi} gas},}\ }\href {\doibase 10.1103/physrevlett.109.220402}
  {\bibfield  {journal} {\bibinfo  {journal} {Phys. Rev. Lett.}\ }\textbf
  {\bibinfo {volume} {109}},\ \bibinfo {pages} {220402} (\bibinfo {year}
  {2012})}\BibitemShut {NoStop}%
\bibitem [{\citenamefont {Hoinka}\ \emph {et~al.}(2013)\citenamefont {Hoinka},
  \citenamefont {Lingham}, \citenamefont {Fenech}, \citenamefont {Hu},
  \citenamefont {Vale}, \citenamefont {Drut},\ and\ \citenamefont
  {Gandolfi}}]{Hoinka2013PRL}%
  \BibitemOpen
  \bibfield  {author} {\bibinfo {author} {\bibfnamefont {S.}~\bibnamefont
  {Hoinka}}, \bibinfo {author} {\bibfnamefont {M.}~\bibnamefont {Lingham}},
  \bibinfo {author} {\bibfnamefont {K.}~\bibnamefont {Fenech}}, \bibinfo
  {author} {\bibfnamefont {H.}~\bibnamefont {Hu}}, \bibinfo {author}
  {\bibfnamefont {C.~J.}\ \bibnamefont {Vale}}, \bibinfo {author}
  {\bibfnamefont {J.~E.}\ \bibnamefont {Drut}}, \ and\ \bibinfo {author}
  {\bibfnamefont {S.}~\bibnamefont {Gandolfi}},\ }\bibfield  {title} {\enquote
  {\bibinfo {title} {Precise determination of the structure factor and contact
  in a unitary fermi gas},}\ }\href {\doibase 10.1103/physrevlett.110.055305}
  {\bibfield  {journal} {\bibinfo  {journal} {Phys. Rev. Lett.}\ }\textbf
  {\bibinfo {volume} {110}},\ \bibinfo {pages} {055305} (\bibinfo {year}
  {2013})}\BibitemShut {NoStop}%
\bibitem [{\citenamefont {Mukherjee}\ \emph {et~al.}(2019)\citenamefont
  {Mukherjee}, \citenamefont {Patel}, \citenamefont {Yan}, \citenamefont
  {Fletcher}, \citenamefont {Struck},\ and\ \citenamefont
  {Zwierlein}}]{Mukherjee2019PRL}%
  \BibitemOpen
  \bibfield  {author} {\bibinfo {author} {\bibfnamefont {B.}~\bibnamefont
  {Mukherjee}}, \bibinfo {author} {\bibfnamefont {P.~B.}\ \bibnamefont
  {Patel}}, \bibinfo {author} {\bibfnamefont {Z.}~\bibnamefont {Yan}}, \bibinfo
  {author} {\bibfnamefont {R.~J.}\ \bibnamefont {Fletcher}}, \bibinfo {author}
  {\bibfnamefont {J.}~\bibnamefont {Struck}}, \ and\ \bibinfo {author}
  {\bibfnamefont {M.~W.}\ \bibnamefont {Zwierlein}},\ }\bibfield  {title}
  {\enquote {\bibinfo {title} {Spectral response and contact of the unitary
  fermi gas},}\ }\href {\doibase 10.1103/physrevlett.122.203402} {\bibfield
  {journal} {\bibinfo  {journal} {Phys. Rev. Lett.}\ }\textbf {\bibinfo
  {volume} {122}},\ \bibinfo {pages} {203402} (\bibinfo {year}
  {2019})}\BibitemShut {NoStop}%
\bibitem [{\citenamefont {Partridge}\ \emph {et~al.}(2005)\citenamefont
  {Partridge}, \citenamefont {Strecker}, \citenamefont {Kamar}, \citenamefont
  {Jack},\ and\ \citenamefont {Hulet}}]{Partridge2005PRL}%
  \BibitemOpen
  \bibfield  {author} {\bibinfo {author} {\bibfnamefont {G.~B.}\ \bibnamefont
  {Partridge}}, \bibinfo {author} {\bibfnamefont {K.~E.}\ \bibnamefont
  {Strecker}}, \bibinfo {author} {\bibfnamefont {R.~I.}\ \bibnamefont {Kamar}},
  \bibinfo {author} {\bibfnamefont {M.~W.}\ \bibnamefont {Jack}}, \ and\
  \bibinfo {author} {\bibfnamefont {R.~G.}\ \bibnamefont {Hulet}},\ }\bibfield
  {title} {\enquote {\bibinfo {title} {Molecular probe of pairing in the
  {BEC}--{BCS} crossover},}\ }\href {\doibase 10.1103/physrevlett.95.020404}
  {\bibfield  {journal} {\bibinfo  {journal} {Phys. Rev. Lett.}\ }\textbf
  {\bibinfo {volume} {95}},\ \bibinfo {pages} {020404} (\bibinfo {year}
  {2005})}\BibitemShut {NoStop}%
\bibitem [{\citenamefont {Chin}\ \emph {et~al.}(2010)\citenamefont {Chin},
  \citenamefont {Grimm}, \citenamefont {Julienne},\ and\ \citenamefont
  {Tiesinga}}]{Chin2010RMP}%
  \BibitemOpen
  \bibfield  {author} {\bibinfo {author} {\bibfnamefont {C.}~\bibnamefont
  {Chin}}, \bibinfo {author} {\bibfnamefont {R.}~\bibnamefont {Grimm}},
  \bibinfo {author} {\bibfnamefont {P.}~\bibnamefont {Julienne}}, \ and\
  \bibinfo {author} {\bibfnamefont {E.}~\bibnamefont {Tiesinga}},\ }\bibfield
  {title} {\enquote {\bibinfo {title} {Feshbach resonances in ultracold
  gases},}\ }\href {\doibase 10.1103/revmodphys.82.1225} {\bibfield  {journal}
  {\bibinfo  {journal} {Rev. Mod. Phys.}\ }\textbf {\bibinfo {volume} {82}},\
  \bibinfo {pages} {1225--1286} (\bibinfo {year} {2010})}\BibitemShut {NoStop}%
\bibitem [{\citenamefont {Wu}\ \emph {et~al.}(2012)\citenamefont {Wu},
  \citenamefont {Park}, \citenamefont {Ahmadi}, \citenamefont {Will},\ and\
  \citenamefont {Zwierlein}}]{Wu2012PRL}%
  \BibitemOpen
  \bibfield  {author} {\bibinfo {author} {\bibfnamefont {C.~H.}\ \bibnamefont
  {Wu}}, \bibinfo {author} {\bibfnamefont {J.~W.}\ \bibnamefont {Park}},
  \bibinfo {author} {\bibfnamefont {P.}~\bibnamefont {Ahmadi}}, \bibinfo
  {author} {\bibfnamefont {S.}~\bibnamefont {Will}}, \ and\ \bibinfo {author}
  {\bibfnamefont {M.~W.}\ \bibnamefont {Zwierlein}},\ }\bibfield  {title}
  {\enquote {\bibinfo {title} {Ultracold fermionic {Feshbach} molecules of
  $^{23}${Na}$^{40}${K}},}\ }\href {\doibase 10.1103/physrevlett.109.085301}
  {\bibfield  {journal} {\bibinfo  {journal} {Phys. Rev. Lett.}\ }\textbf
  {\bibinfo {volume} {109}},\ \bibinfo {pages} {085301} (\bibinfo {year}
  {2012})}\BibitemShut {NoStop}%
\bibitem [{\citenamefont {Zhang}\ and\ \citenamefont
  {Leggett}(2009)}]{ZhangSZ2009}%
  \BibitemOpen
  \bibfield  {author} {\bibinfo {author} {\bibfnamefont {S.}~\bibnamefont
  {Zhang}}\ and\ \bibinfo {author} {\bibfnamefont {A.~J.}\ \bibnamefont
  {Leggett}},\ }\bibfield  {title} {\enquote {\bibinfo {title} {Universal
  properties of the ultracold {F}ermi gas},}\ }\href {\doibase
  10.1103/physreva.79.023601} {\bibfield  {journal} {\bibinfo  {journal} {Phys.
  Rev. A}\ }\textbf {\bibinfo {volume} {79}},\ \bibinfo {pages} {023601}
  (\bibinfo {year} {2009})}\BibitemShut {NoStop}%
\bibitem [{\citenamefont {Yao}\ \emph {et~al.}(2016)\citenamefont {Yao},
  \citenamefont {Chen}, \citenamefont {Wu}, \citenamefont {Liu}, \citenamefont
  {Wang}, \citenamefont {Jiang}, \citenamefont {Deng}, \citenamefont {Chen},\
  and\ \citenamefont {Pan}}]{Yao2016PRL}%
  \BibitemOpen
  \bibfield  {author} {\bibinfo {author} {\bibfnamefont {X.-C.}\ \bibnamefont
  {Yao}}, \bibinfo {author} {\bibfnamefont {H.-Z.}\ \bibnamefont {Chen}},
  \bibinfo {author} {\bibfnamefont {Y.-P.}\ \bibnamefont {Wu}}, \bibinfo
  {author} {\bibfnamefont {X.-P.}\ \bibnamefont {Liu}}, \bibinfo {author}
  {\bibfnamefont {X.-Q.}\ \bibnamefont {Wang}}, \bibinfo {author}
  {\bibfnamefont {X.}~\bibnamefont {Jiang}}, \bibinfo {author} {\bibfnamefont
  {Y.}~\bibnamefont {Deng}}, \bibinfo {author} {\bibfnamefont {Y.-A.}\
  \bibnamefont {Chen}}, \ and\ \bibinfo {author} {\bibfnamefont {J.-W.}\
  \bibnamefont {Pan}},\ }\bibfield  {title} {\enquote {\bibinfo {title}
  {Observation of coupled vortex lattices in a mass-imbalance {Bose} and
  {Fermi} superfluid mixture},}\ }\href {\doibase
  10.1103/physrevlett.117.145301} {\bibfield  {journal} {\bibinfo  {journal}
  {Phys. Rev. Lett.}\ }\textbf {\bibinfo {volume} {117}},\ \bibinfo {pages}
  {145301} (\bibinfo {year} {2016})}\BibitemShut {NoStop}%
\bibitem [{\citenamefont {Wu}\ \emph {et~al.}(2017)\citenamefont {Wu},
  \citenamefont {Yao}, \citenamefont {Chen}, \citenamefont {Liu}, \citenamefont
  {Wang}, \citenamefont {Chen},\ and\ \citenamefont {Pan}}]{Wu2017JPB}%
  \BibitemOpen
  \bibfield  {author} {\bibinfo {author} {\bibfnamefont {Y.-P.}\ \bibnamefont
  {Wu}}, \bibinfo {author} {\bibfnamefont {X.-C.}\ \bibnamefont {Yao}},
  \bibinfo {author} {\bibfnamefont {H.-Z.}\ \bibnamefont {Chen}}, \bibinfo
  {author} {\bibfnamefont {X.-P.}\ \bibnamefont {Liu}}, \bibinfo {author}
  {\bibfnamefont {X.-Q.}\ \bibnamefont {Wang}}, \bibinfo {author}
  {\bibfnamefont {Y.-A.}\ \bibnamefont {Chen}}, \ and\ \bibinfo {author}
  {\bibfnamefont {J.-W.}\ \bibnamefont {Pan}},\ }\bibfield  {title} {\enquote
  {\bibinfo {title} {A quantum degenerate {Bose}-{Fermi} mixture of $^{41}${K}
  and $^6${Li}},}\ }\href {http://stacks.iop.org/0953-4075/50/i=9/a=094001}
  {\bibfield  {journal} {\bibinfo  {journal} {J. Phys. B: At., Mol. Opt.
  Phys.}\ }\textbf {\bibinfo {volume} {50}},\ \bibinfo {pages} {094001}
  (\bibinfo {year} {2017})}\BibitemShut {NoStop}%
\bibitem [{\citenamefont {Ravensbergen}\ \emph {et~al.}(2018)\citenamefont
  {Ravensbergen}, \citenamefont {Corre}, \citenamefont {Soave}, \citenamefont
  {Kreyer}, \citenamefont {Tzanova}, \citenamefont {Kirilov},\ and\
  \citenamefont {Grimm}}]{Ravensbergen2018PRL}%
  \BibitemOpen
  \bibfield  {author} {\bibinfo {author} {\bibfnamefont {C.}~\bibnamefont
  {Ravensbergen}}, \bibinfo {author} {\bibfnamefont {V.}~\bibnamefont {Corre}},
  \bibinfo {author} {\bibfnamefont {E.}~\bibnamefont {Soave}}, \bibinfo
  {author} {\bibfnamefont {M.}~\bibnamefont {Kreyer}}, \bibinfo {author}
  {\bibfnamefont {S.}~\bibnamefont {Tzanova}}, \bibinfo {author} {\bibfnamefont
  {E.}~\bibnamefont {Kirilov}}, \ and\ \bibinfo {author} {\bibfnamefont
  {R.}~\bibnamefont {Grimm}},\ }\bibfield  {title} {\enquote {\bibinfo {title}
  {Accurate determination of the dynamical polarizability of dysprosium},}\
  }\href {\doibase 10.1103/physrevlett.120.223001} {\bibfield  {journal}
  {\bibinfo  {journal} {Phys. Rev. Lett.}\ }\textbf {\bibinfo {volume} {120}},\
  \bibinfo {pages} {223001} (\bibinfo {year} {2018})}\BibitemShut {NoStop}%
\bibitem [{\citenamefont {Stajic}\ \emph {et~al.}(2004)\citenamefont {Stajic},
  \citenamefont {Milstein}, \citenamefont {Chen}, \citenamefont {Chiofalo},
  \citenamefont {Holland},\ and\ \citenamefont {Levin}}]{JS2}%
  \BibitemOpen
  \bibfield  {author} {\bibinfo {author} {\bibfnamefont {J.}~\bibnamefont
  {Stajic}}, \bibinfo {author} {\bibfnamefont {J.~N.}\ \bibnamefont
  {Milstein}}, \bibinfo {author} {\bibfnamefont {Q~J}\ \bibnamefont {Chen}},
  \bibinfo {author} {\bibfnamefont {M.~L.}\ \bibnamefont {Chiofalo}}, \bibinfo
  {author} {\bibfnamefont {M.~J.}\ \bibnamefont {Holland}}, \ and\ \bibinfo
  {author} {\bibfnamefont {K.}~\bibnamefont {Levin}},\ }\bibfield  {title}
  {\enquote {\bibinfo {title} {The nature of superfluidity in ultracold {F}ermi
  gases near {F}eshbach resonances},}\ }\href {\doibase
  10.1103/PhysRevA.69.063610} {\bibfield  {journal} {\bibinfo  {journal} {Phys.
  Rev. A}\ }\textbf {\bibinfo {volume} {69}},\ \bibinfo {pages} {063610}
  (\bibinfo {year} {2004})}\BibitemShut {NoStop}%
\bibitem [{\citenamefont {Z\"urn}\ \emph {et~al.}(2013)\citenamefont {Z\"urn},
  \citenamefont {Lompe}, \citenamefont {Wenz}, \citenamefont {Jochim},
  \citenamefont {Julienne},\ and\ \citenamefont {Hutson}}]{Zuern2013PRL}%
  \BibitemOpen
  \bibfield  {author} {\bibinfo {author} {\bibfnamefont {G.}~\bibnamefont
  {Z\"urn}}, \bibinfo {author} {\bibfnamefont {T.}~\bibnamefont {Lompe}},
  \bibinfo {author} {\bibfnamefont {A.~N.}\ \bibnamefont {Wenz}}, \bibinfo
  {author} {\bibfnamefont {S.}~\bibnamefont {Jochim}}, \bibinfo {author}
  {\bibfnamefont {P.~S.}\ \bibnamefont {Julienne}}, \ and\ \bibinfo {author}
  {\bibfnamefont {J.~M.}\ \bibnamefont {Hutson}},\ }\bibfield  {title}
  {\enquote {\bibinfo {title} {Precise characterization of $^6${Li} {Feshbach}
  resonances using trap-sideband-resolved {RF} spectroscopy of weakly bound
  molecules},}\ }\href {\doibase 10.1103/physrevlett.110.135301} {\bibfield
  {journal} {\bibinfo  {journal} {Phys. Rev. Lett.}\ }\textbf {\bibinfo
  {volume} {110}},\ \bibinfo {pages} {135301} (\bibinfo {year}
  {2013})}\BibitemShut {NoStop}%
\bibitem [{\citenamefont {Sanner}\ \emph {et~al.}(2012)\citenamefont {Sanner},
  \citenamefont {Su}, \citenamefont {Huang}, \citenamefont {Keshet},
  \citenamefont {Gillen},\ and\ \citenamefont {Ketterle}}]{Sanner2012PRL}%
  \BibitemOpen
  \bibfield  {author} {\bibinfo {author} {\bibfnamefont {C.}~\bibnamefont
  {Sanner}}, \bibinfo {author} {\bibfnamefont {E.~J.}\ \bibnamefont {Su}},
  \bibinfo {author} {\bibfnamefont {W.}~\bibnamefont {Huang}}, \bibinfo
  {author} {\bibfnamefont {A.}~\bibnamefont {Keshet}}, \bibinfo {author}
  {\bibfnamefont {J.}~\bibnamefont {Gillen}}, \ and\ \bibinfo {author}
  {\bibfnamefont {W.}~\bibnamefont {Ketterle}},\ }\bibfield  {title} {\enquote
  {\bibinfo {title} {Correlations and pair formation in a repulsively
  interacting {Fermi} gas},}\ }\href {\doibase 10.1103/physrevlett.108.240404}
  {\bibfield  {journal} {\bibinfo  {journal} {Phys. Rev. Lett.}\ }\textbf
  {\bibinfo {volume} {108}},\ \bibinfo {pages} {240404} (\bibinfo {year}
  {2012})}\BibitemShut {NoStop}%
\bibitem [{\citenamefont {Ku}\ \emph {et~al.}(2012)\citenamefont {Ku},
  \citenamefont {Sommer}, \citenamefont {Cheuk},\ and\ \citenamefont
  {Zwierlein}}]{Zwierlein2011}%
  \BibitemOpen
  \bibfield  {author} {\bibinfo {author} {\bibfnamefont {M.~J.~H.}\
  \bibnamefont {Ku}}, \bibinfo {author} {\bibfnamefont {A.~T.}\ \bibnamefont
  {Sommer}}, \bibinfo {author} {\bibfnamefont {L.~W.}\ \bibnamefont {Cheuk}}, \
  and\ \bibinfo {author} {\bibfnamefont {M.~W.}\ \bibnamefont {Zwierlein}},\
  }\bibfield  {title} {\enquote {\bibinfo {title} {Revealing the superfluid
  lambda transition in the universal thermodynamics of a unitary {Fermi}
  gas},}\ }\href {\doibase 10.1126/science.1214987} {\bibfield  {journal}
  {\bibinfo  {journal} {Science}\ }\textbf {\bibinfo {volume} {335}},\ \bibinfo
  {pages} {563–567} (\bibinfo {year} {2012})}\BibitemShut {NoStop}%
\bibitem [{\citenamefont {Ko}\ \emph {et~al.}(2019)\citenamefont {Ko},
  \citenamefont {Park},\ and\ \citenamefont {Shin}}]{Ko2019NP}%
  \BibitemOpen
  \bibfield  {author} {\bibinfo {author} {\bibfnamefont {B.}~\bibnamefont
  {Ko}}, \bibinfo {author} {\bibfnamefont {J.~W.}\ \bibnamefont {Park}}, \ and\
  \bibinfo {author} {\bibfnamefont {Y.}~\bibnamefont {Shin}},\ }\bibfield
  {title} {\enquote {\bibinfo {title} {Kibble{\textendash}zurek universality in
  a strongly interacting fermi superfluid},}\ }\href {\doibase
  10.1038/s41567-019-0650-1} {\bibfield  {journal} {\bibinfo  {journal} {Nat.
  Phys.}\ }\textbf {\bibinfo {volume} {15}},\ \bibinfo {pages} {1227--1231}
  (\bibinfo {year} {2019})}\BibitemShut {NoStop}%
\bibitem [{\citenamefont {Heiselberg}(2001)}]{Heiselberg2001PRA}%
  \BibitemOpen
  \bibfield  {author} {\bibinfo {author} {\bibfnamefont {H.}~\bibnamefont
  {Heiselberg}},\ }\bibfield  {title} {\enquote {\bibinfo {title} {Fermi
  systems with long scattering lengths},}\ }\href {\doibase
  10.1103/PhysRevA.63.043606} {\bibfield  {journal} {\bibinfo  {journal}
  {\pra}\ }\textbf {\bibinfo {volume} {63}},\ \bibinfo {pages} {043606}
  (\bibinfo {year} {2001})}\BibitemShut {NoStop}%
\bibitem [{\citenamefont {Carlson}\ \emph {et~al.}(2003)\citenamefont
  {Carlson}, \citenamefont {Chang}, \citenamefont {Pandharipande},\ and\
  \citenamefont {Schmidt}}]{Carlson2003PRL}%
  \BibitemOpen
  \bibfield  {author} {\bibinfo {author} {\bibfnamefont {J.}~\bibnamefont
  {Carlson}}, \bibinfo {author} {\bibfnamefont {S.-Y.}\ \bibnamefont {Chang}},
  \bibinfo {author} {\bibfnamefont {V.~R.}\ \bibnamefont {Pandharipande}}, \
  and\ \bibinfo {author} {\bibfnamefont {K.~E.}\ \bibnamefont {Schmidt}},\
  }\bibfield  {title} {\enquote {\bibinfo {title} {Superfluid fermi gases with
  large scattering length},}\ }\href {\doibase 10.1103/PhysRevLett.91.050401}
  {\bibfield  {journal} {\bibinfo  {journal} {\prl}\ }\textbf {\bibinfo
  {volume} {91}},\ \bibinfo {pages} {050401} (\bibinfo {year}
  {2003})}\BibitemShut {NoStop}%
\bibitem [{\citenamefont {Ho}(2004)}]{Ho2004PRL}%
  \BibitemOpen
  \bibfield  {author} {\bibinfo {author} {\bibfnamefont {Tin-Lun}\ \bibnamefont
  {Ho}},\ }\bibfield  {title} {\enquote {\bibinfo {title} {Universal
  thermodynamics of degenerate quantum gases in the unitarity limit},}\ }\href
  {\doibase 10.1103/physrevlett.92.090402} {\bibfield  {journal} {\bibinfo
  {journal} {Phys. Rev. Lett.}\ }\textbf {\bibinfo {volume} {92}},\ \bibinfo
  {pages} {090402} (\bibinfo {year} {2004})}\BibitemShut {NoStop}%
\end{thebibliography}%

\end{document}